\documentclass[8pt, twocolumn, a4paper]{article}

\usepackage[utf8]{inputenc} 
\usepackage[T1]{fontenc}    

\usepackage{siunitx}        
\usepackage{url}            
\usepackage{booktabs}       
\usepackage{amsfonts}       
\usepackage{amsmath}        
\usepackage{amssymb}        
\usepackage{mathtools}      
\usepackage{graphicx}       
\usepackage{caption}        
\usepackage{subcaption}     
\usepackage{makecell}       
\usepackage{multirow, bigdelim} 
\usepackage[linesnumbered]{algorithm2e} 
\usepackage[percent]{overpic} 
\usepackage{xcolor}         

\usepackage[numbers, sort&compress]{natbib} 
\usepackage{authblk}        
\usepackage{hyperref}       
\usepackage{cleveref}       
\usepackage{placeins}

\hypersetup{
    colorlinks=true,
    linkcolor=blue,
    filecolor=magenta,
    urlcolor=cyan,
    citecolor=blue, 
    pdftitle={Automatic Raman Measurements in a High-Throughput Bioprocess Development Lab},
    pdfauthor={Christoph Lange, Simon Seidel, Madeline Altmann, Daniel Stors, Annina Kemmer, Linda Cai, Stefan Born, Peter Neubauer, M. Nicolas Cruz Bournazou},
    pdfsubject={Biotechnology and Bioengineering},
    pdfkeywords={Raman Spectroscopy, Automation, High-Throughput Bioprocessing},
}

\title{Automatic Raman Measurements in a High-Throughput Bioprocess Development Lab}

\author[1]{Christoph Lange\thanks{Equally contributing authors. Corresponding authors: christoph.lange@tu-berlin.de, simon.seidel@tu-berlin.de}}
\author[1]{Simon Seidel$^*$}
\author[1]{Madeline Altmann}
\author[1]{Daniel Stors}
\author[1]{Annina Kemmer}
\author[1]{Linda Cai}
\author[2]{Stefan Born}
\author[1]{Peter Neubauer}
\author[1]{M. Nicolas Cruz Bournazou}

\affil[1]{Technische Universität Berlin, Chair of Bioprocess Engineering, Straße des 17. Juni 135, 10623 Berlin, Germany}
\affil[2]{Technische Universität Berlin, Orientierungsstudium MINTgrün, Straße des 17. Juni 135, 10623 Berlin, Germany}

\date{} 

\begin{document}

\maketitle 

\begin{abstract}
This study presents a collection of physical devices and software services that fully automate Raman spectra measurements for liquid samples within a robotic facility. This method is applicable to various fields, with demonstrated efficacy in biotechnology, where Raman spectroscopy monitors substrates, metabolites, and product-related concentrations.
Our system specifically measures \qty{50}{\micro \litre} samples using a liquid handling robot capable of taking $8$ samples simultaneously. We record multiple Raman spectra for \qty{10}{\second} each. Furthermore, our system takes around \qty{20}{\second} for sample handling, cleaning, and preparation of the next measurement. All spectra and metadata are stored in a database and we use a machine learning model to estimate concentrations from the spectra.
This automated approach enables gathering spectra for various applications under uniform conditions in high-throughput fermentation processes, calibration procedures, and offline evaluations. This allows data to be combined to train sophisticated machine learning models with improved generalization. Consequently, we can develop accurate models more quickly for new applications by reusing data from prior applications, thereby reducing the need for extensive calibration data.

\vspace{1em} 
\end{abstract}

\section{Introduction}
High-throughput cultivation systems are crucial for advancing modern bioprocess development \cite{neubauer_consistent_2013,long_development_2014,kopp_high-throughput_2023}. In particular the implementation of automated systems, non-invasive sensors, and liquid handling technologies enhances process control \cite{hemmerich_microbioreactor_2018,teworte_recent_2022}. Automation enables the conduction of a larger quantity of parallel fed-batch cultivations with advanced feeding logics, leading to improved reproducibility \cite{qian_fully_2021, anantanawat_high-throughput_2019,seidel_thermal_2024} and elevating the sampling frequency, thereby permitting more precise monitoring of the fermentations.

Within the framework of Process Analytical Technology (PAT), this denotes enhancing the regulation of manufacturing processes whilst safeguarding product quality. In the realm of PAT, Raman spectroscopy has attained significance \cite{lin_raman_2021} owing to its capability to monitor substrates, metabolites, and product concentrations in a noninvasive manner \cite{wang_-line_2023, allakhverdiev_raman_2022}. Its proficiency in delivering swift and comprehensive molecular information within a single spectrum \cite{lourenco_bioreactor_2012} makes Raman spectroscopy an efficient PAT, for instance \cite{nemcova_use_2021}.

Although the precision of information regarding process parameters obtained from Raman spectra may not match the level achieved through techniques such as High-performance Liquid Chromatography \cite{muller_bioprocess_2023}, it provides the benefit of necessitating only minimal sample volumes and facilitates swift information retrieval for numerous quantities of interest, particularly when combined with machine learning models \cite{esmonde-white_raman_2017, luo_deep_2022, rowlandjones_atline_2019, goldrick_high-throughput_2020}.
The use of Raman spectroscopy for monitoring Chinese hamster ovary (CHO) cells is well-documented in the literature \cite{rowlandjones_atline_2019,webster_development_2018,yousefi-darani_generic_2022}. Using CHO cells, several authors have demonstrated the feasibility of Raman spectroscopy for viable cell density measurements with cell counts of up to  1 \( \cdot \) 10  \textsuperscript{8} cells/\unit{\milli \litre} \cite{schwarz_monitoring_2022}. Additionally, there are systems designed to measure Raman spectra in a high-throughput context at the \qty{15}{\milli \litre} \cite{rowlandjones_spectroscopy_2021} and \qty{250}{\milli \litre} \cite{graf_automated_2022} scales. However, these systems are limited to measuring only one sample at a time, with a recording duration of \qtyrange{1}{5}{\min} per spectrum. Furthermore, both hardware and software are proprietary, which hinders modifications, integration into existing systems and deploying potent machine learning models like neural networks for predicting concentrations. Furthermore, measuring Raman spectra in high optical density bacterial fermentations requires flexible modifications for sample preparation to enhance signal intensity \cite{rodriguez_recent_2023,goldrick_high-throughput_2020}.

Our objective is to develop a system for the automated measurement of Raman spectra that seamlessly integrates with any high-throughput cultivation platform. We showcase the implementation for $48$ minibioreactors \cite{kopp_high-throughput_2023}. Within this framework, in situ measurement of Raman spectra is rendered impractical due to the prohibitive cost of the numerous spectrometers required. Furthermore, in-line recording using a flow cell connected to all bioreactors presents challenges, as the reactors are single-use and lack compatible mounting interfaces. Consequently, we have opted to conduct measurements in-line to expedite measurement times \cite{hirsch_inline_2019}, minimize human error, and allow flexible control of all interfaces to our auxiliary devices.
Moreover, we would like the system to be consistent with offline analytics and across scales, i.e. the spectra do not depend on the characteristics of the reactor in which they were recorded.

%

In this study, we comprehensively detail all elements involved in Raman measurements. Initially, we examine the hardware components, succeeded by an analysis of the software and their integrated functionality. Subsequently, we propose a calibration procedure aimed at enhancing signal quality and show its use in a fermentation of \textit{Escherichia coli}.

\begin{figure*}[htbp]
    \centering
    \begin{minipage}[t]{0.7\textwidth} 
        \vspace{0pt} 
        \begin{minipage}[t]{0.46\textwidth} 
            \centering
            \begin{overpic}[width=\textwidth]{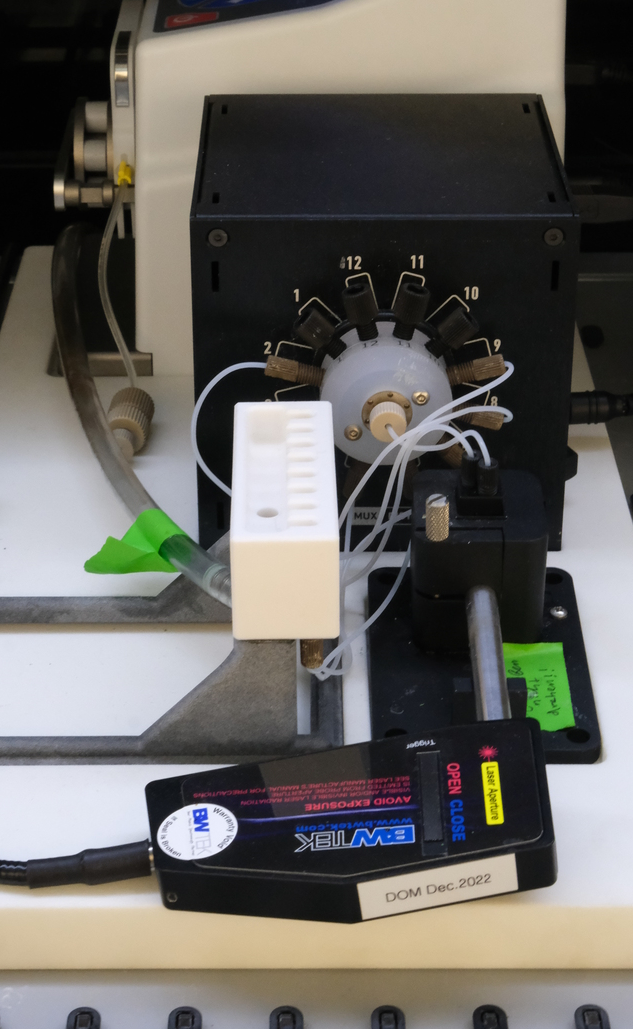}
                \put(1,97){\textbf{\textcolor{white}{a)}}} 
            \end{overpic}
            \label{fig:PhotographRealSetup}
        \end{minipage}%
        \hfill
        \begin{minipage}[t]{0.50\textwidth} 
            \centering
            \begin{overpic}[width=\textwidth]{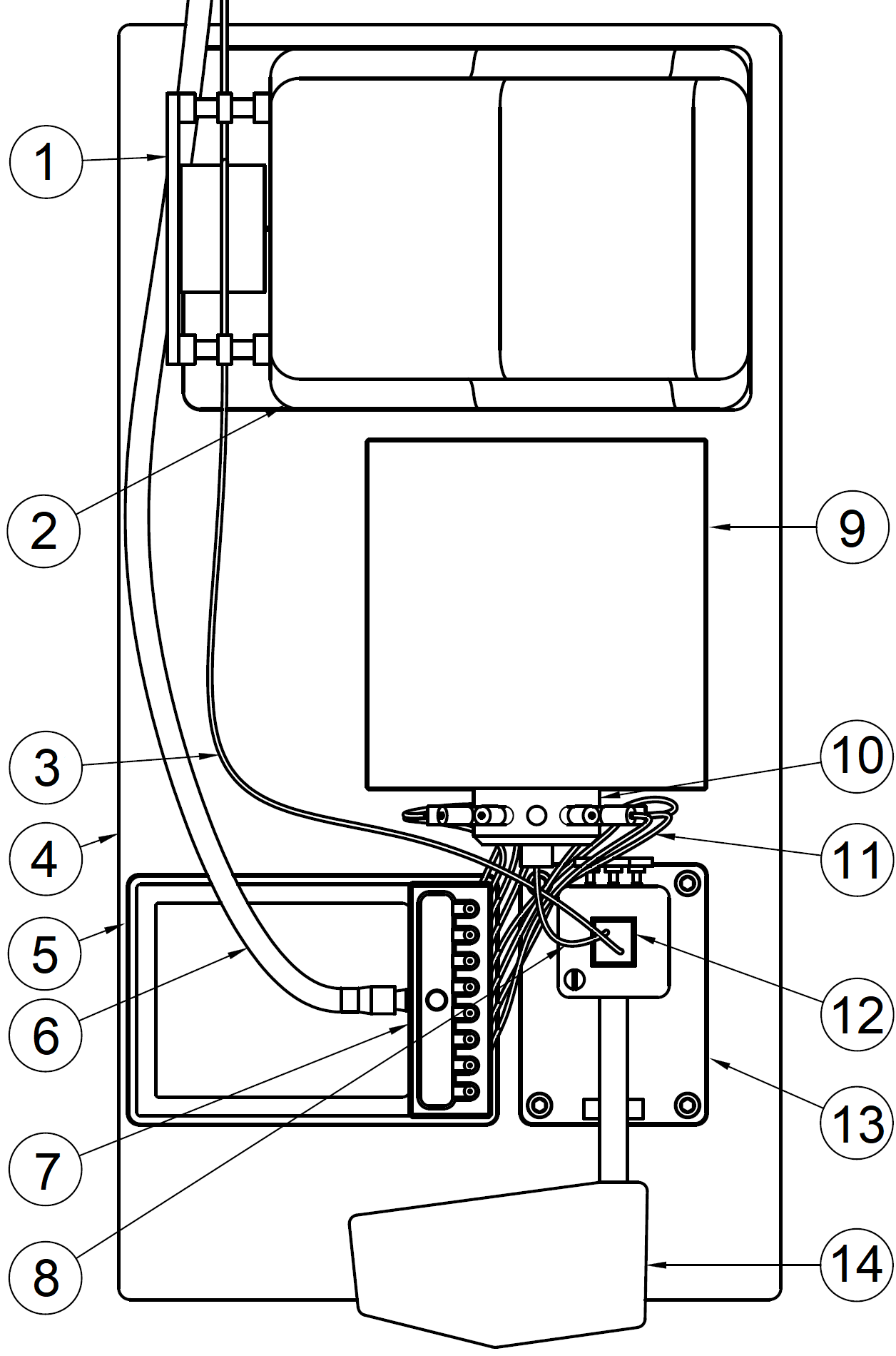}
                \put(5,95){\textbf{b)}} 
            \end{overpic}
            \label{fig:GesamtSchema}
        \end{minipage}
    \end{minipage}
    \begin{minipage}[t]{0.25\textwidth} 
        \vspace{0pt} 
        \begin{subfigure}[t]{\textwidth} 
            \centering
            \begin{overpic}[width=\textwidth]{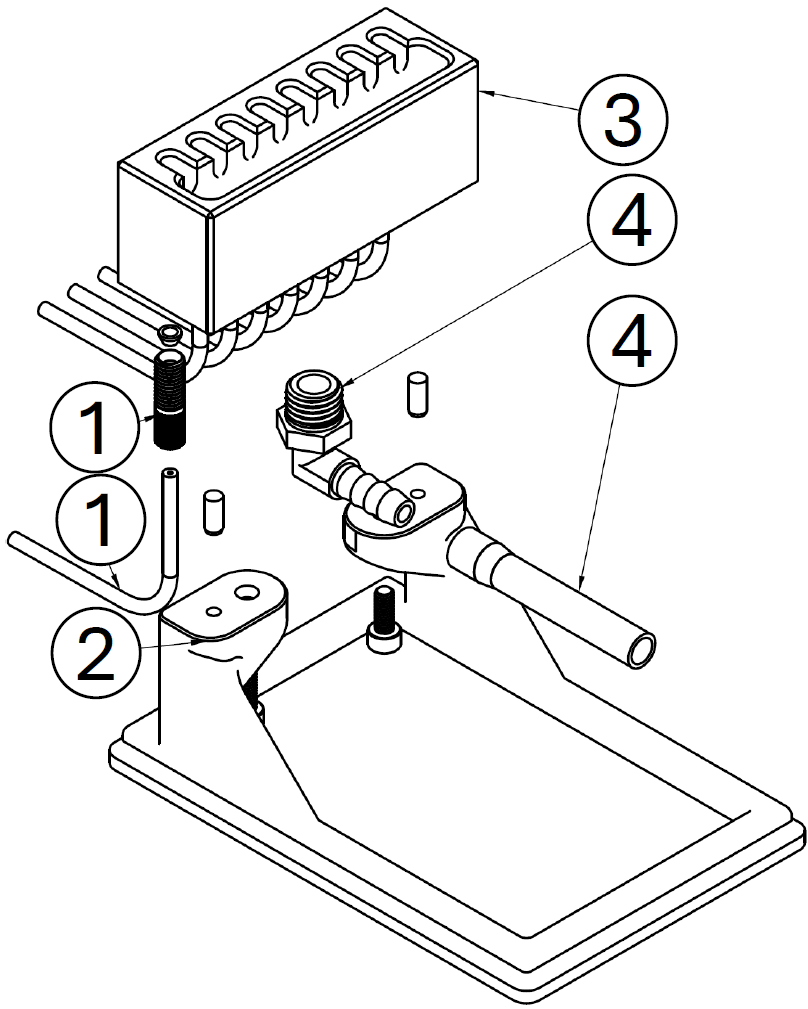}
                \put(5,95){\textbf{c)}}  
            \end{overpic}
            \label{fig:Explosiondrawing}
        \end{subfigure}
        \vspace{1em}
        \hspace{1em}
        \begin{subfigure}[t]{0.4\textwidth} 
            \centering
            \begin{overpic}[width=\textwidth]{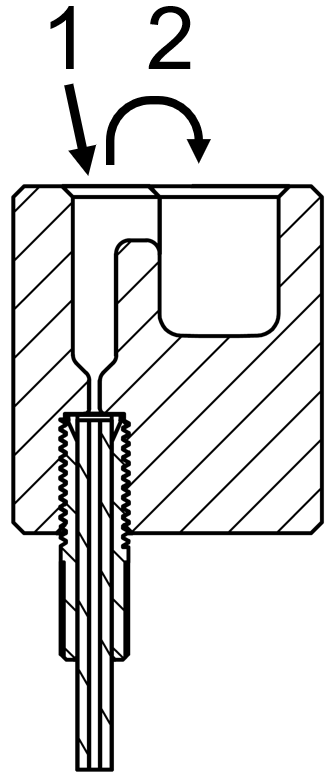}
                \put(-3,95){\textbf{d)}}  
            \end{overpic}
            \label{fig:SamplerSchnittbildV3}   
        \end{subfigure}
    \end{minipage}
    \caption{a) Devices Placement in the Liquid Handling Station:
The integrated setup for automated Raman spectral measurements within a liquid handling robot. It includes a PTFE pipetting interface for up to eight samples, linked via microfluidic tubing to a multiplexer valve (Elveflow, Paris, France) for rapid transport to a flow-through cuvette (Hellma GmbH \& Co. KG, Müllheim, Germany). A Metrohm i-Raman Plus 785 Spectrometer probe is connected to the cuvette holder. b) Schematic Top View of Devices:
The components include: (1) peristaltic pump head, (2) pump case, (3) pumping tube, (4) device platform, (5) sampling interface rack, (6) wastewater tube, (7) sampling interface, (8) tube connecting valve and cuvette, (9) multiplexer valve's case, (10) multiplexer head, (11) microfluidic tubes connecting sampling wells and valve, (12) flow-through cuvette, (13) cuvette holder, and (14) Raman probe. c) Sample Interface: It features eight wells (3) connected to the multiplexer valve via mounted tubes (1). A tube is connected to the overflow of the interface (4) and the whole item is sitting on a stand (2).
d) Transverse Section of the Sampling Device: A cross-sectional view of a sampling well from the CAD model of the pipetting interface. Functions include: (1) introducing the sample into the well via pipetting; and (2) overflow of the sample and cleaning solution, which are pumped backward into the drainage system.
}
\label{fig:Setup}
\end{figure*}


\section{Material \& Methods} \label{mm_section}
An exhaustive exposition of the physical devices is initially presented in Section \ref{devices_section}, subsequently we present more details about the software components in Section \ref{software_section}.

\subsection{Devices} \label{devices_section}
The system, as depicted in Figure \ref{fig:Setup} a) and b), comprises multiple units positioned within a liquid handling station to facilitate unobstructed transitions of samples. The sampling interface (Section \ref{interface_device}) retrieves samples from the liquid handling robot (Section \ref{Liquid_Handling_Robot}). Subsequently, each sample is transferred to the cuvette (Section \ref{cuvette_device}) via a pump (Section \ref{pump_device}). This configuration requires supplementary devices, including a multiplexer (Section \ref{mux_device}), to interconnect each well with the flow-through cuvette, thereby allowing the spectrometer (Section \ref{spectrometer_device}) to conduct the recording process.

\subsubsection{Interface} \label{interface_device}
We designed a chemically inert sampling interface (Figure \ref{fig:Setup} c)) made from polytetrafluoroethylene (PTFE) to accept samples from a liquid handling robot. Each of its 8 wells can hold a volume of \qty{125}{\milli \liter} and the wells are spaced \qty{9}{\milli \meter} apart which is the same distance used for $96$-well microtiter plates. Each well is connected through a PTFE tube with inner diameter of \qty{0.508}{\milli \meter} to the multiplexer (Section \ref{mux_device}) using microfluidic fittings. This setup allows a parallel acceptance of 8 samples (arrow 1 in Figure \ref{fig:Setup} d)) that wait for their sequential spectra measurement. The device can also flush samples into the wastewater container, via the overflow (arrow 2 in Figure \ref{fig:Setup} d)).

\subsubsection{Multiplexer Valve} \label{mux_device}
We link the eight sampling interface wells (refer to Section \ref{interface_device}) to the cuvette (see Section \ref{cuvette_device}) by employing a 12-to-1 bidirectional valve multiplexer (Elveflow, Paris, France). Specifically, we utilize the multiplexer head's eight lower ports (labeled 2 through 9 in Figure \ref{fig:Setup} b)) to minimize the distance between each port and the sampling interface. The multiplexer valve establishes a connection between one external port, which links to the sampling interface wells (discussed in Section \ref{interface_device}), and the central port that connects to the cuvette \ref{cuvette_device} through a \qty{65}{\milli \meter} PTFE tube. All tubes linking the wells to the multiplexer ports measure \qty{125}{\milli \meter} in length with a diameter of \qty{0,508}{\milli \meter}. We optimized them for minimal volumes in the tubes and an elevated pump speed to enhance sample throughput.


\subsubsection{Cuvette and Cuvette Holder} \label{cuvette_device}
Sample measurements are conducted using a flow-through cuvette with standard dimensions (refer to Figure \ref{fig:Setup} b, item 10) within a cuvette holder (model BCR100A, B\&W Tec, USA, see Figure \ref{fig:Setup} b, item 8). This holder enhances signal strength by utilizing a mirror, thereby reducing the time required to record each spectrum. Acting as a flow cell, the cuvette has a capacity of \qty{18}{\micro \litre} and features an optical path length of \qty{10}{\milli \meter} (manufactured by Hellma GmbH \& Co. KG, Article No. 1787128510-40).

\subsubsection{Spectrometer} \label{spectrometer_device}
We employ a Metrohm i-Raman Plus 785 Spectrometer (Metrohm, Herisau, Switzerland) featuring an excitation wavelength of \qty{785}{\nano \meter}, optimizing the balance between signal intensity and fluorescence \cite{McCreery_raman_2000} for biotechnological applications. The spectra encompass $2048$ dimensions within the range of \qtyrange{65}{3350} {\per \centi \meter}, and the laser power extends up to \qty{455}{\milli \watt}, enabling swift data acquisition. The BAC102 probe (B\&W Tec, USA, depicted in Figure \ref{fig:Setup} b) item 14) is utilized.
\subsubsection{Pump} \label{pump_device}
For the purpose of transferring the sample from the sampling interface to the flow cell, a peristaltic pump was employed. Peristaltic pumps are recognized for delivering precise and reproducible pumping characteristics at high speeds in microfluidic systems \cite{tamadon_miniaturized_2019}. Furthermore, these pumps exert a gentle influence on fluids, thereby reducing mixing between the sample designated for measurement and the washing solution \cite{Klespitz_peristaltic_2014}. In this specific process, a Masterflex Ismatec Peristaltic Pump, REGLO ICC (Avantor, Inc., Radnor, Pennsylvania, USA), was utilized.

\subsubsection{Liquid Handling Robot}  \label{Liquid_Handling_Robot}
The system illustrated in Figure \ref{fig:Setup} is incorporated into a liquid handling station, namely a Tecan EVO 200 (Tecan Group, Männedorf, Switzerland), which facilitates automatic measurements during calibration, cultivation, or when utilized as an offline analyzer. This liquid handling robot is equipped with an arm featuring eight steel needles, enabling the transfer of samples into the eight wells of the sampling interface (Section \ref{interface_device}). In addition, we use a mini bioreactor system (48 BioReactor; 2mag AG) on the same liquid handler.

\subsection{Software Components} \label{software_section}

\begin{figure*}[!ht] 
    \centering
    \begin{subfigure}[t]{\textwidth}
    \includegraphics[width=0.9\linewidth]{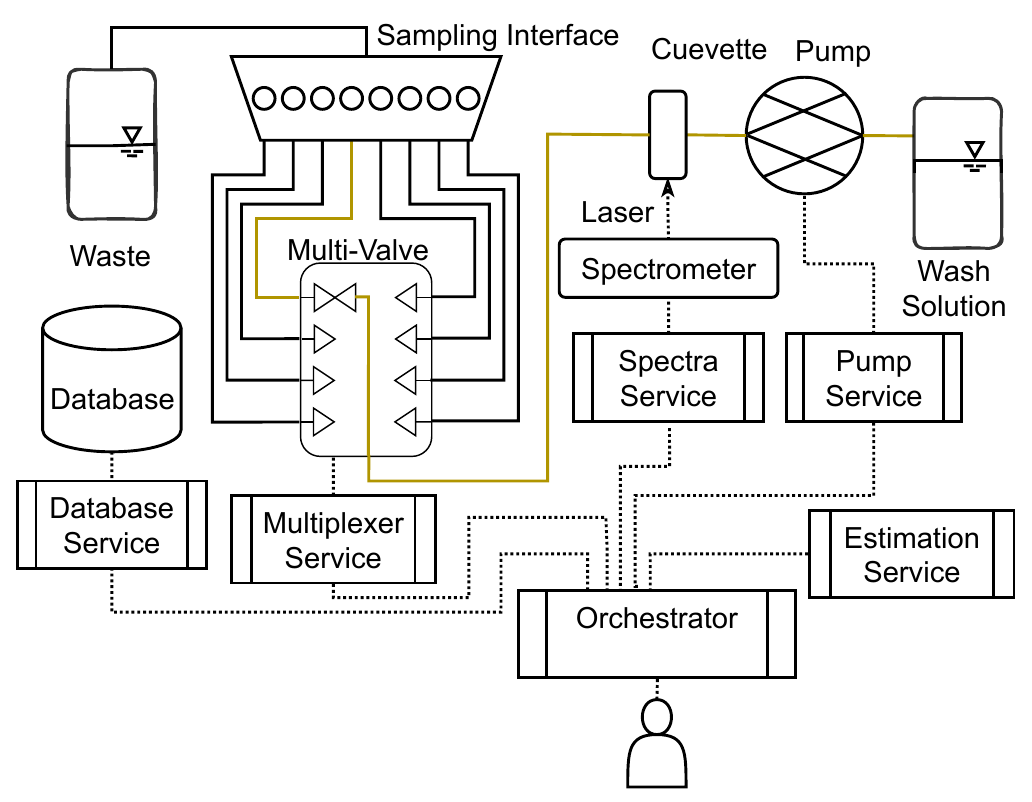}
    \caption{\textit{Interaction of Components needed for a Raman Measurement}: Here we depict all the physical components as well as the software elements (black boxes with two lines on each side) that are involved in an automatic Raman measurement. The black lines schematically represent the connection between the devices via tubes.  Dashed lines indicate connections realized digitally.}
    \label{fig:overview}
    \end{subfigure}
    \hfill
    \begin{subfigure}[t]{0.32\textwidth}
    \includegraphics[width=\textwidth]{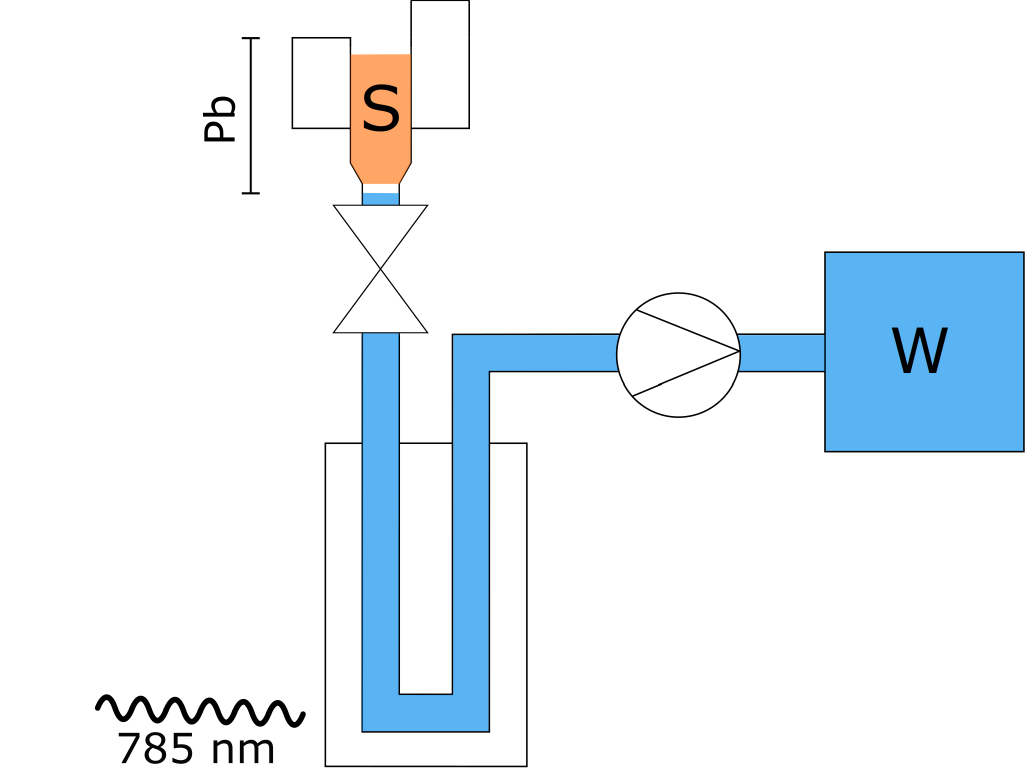}
    \caption{Sample (S) in the interface waiting for measurement}
    \label{fig:sample_waiting}
    \end{subfigure}
    \hfill
    \begin{subfigure}[t]{0.32\textwidth}
    \includegraphics[width=\textwidth]{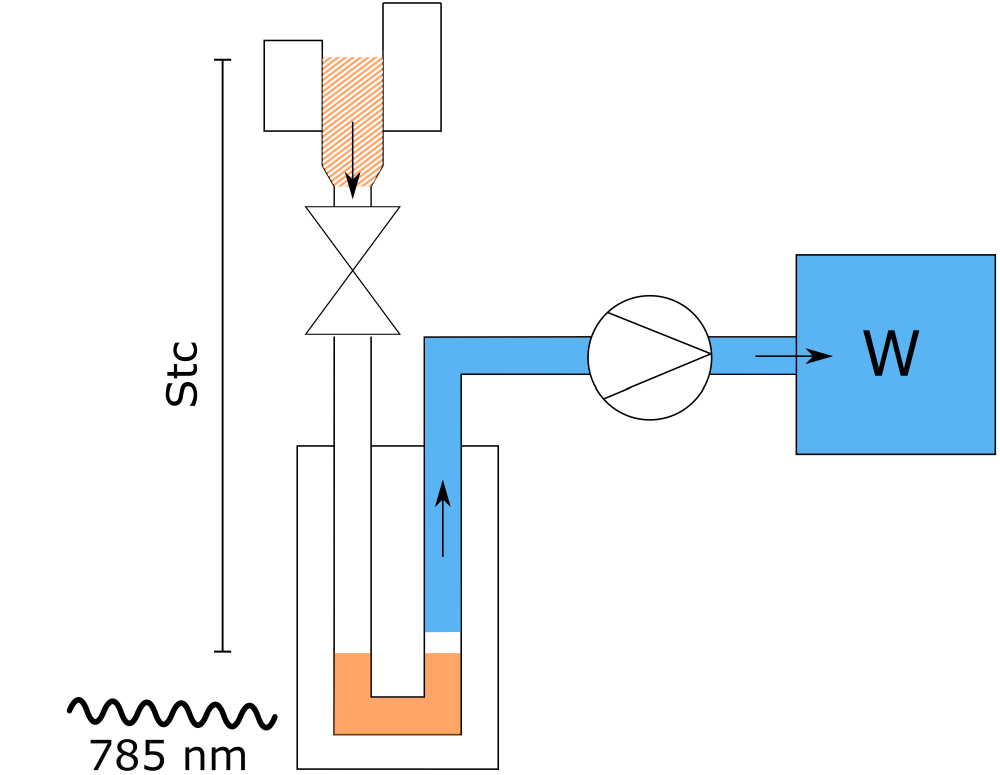}
    \caption{Measuring sample in cuvette}
    \label{fig:sample_measurement}
    \end{subfigure}
   \begin{subfigure}[t]{0.32\textwidth}
    \includegraphics[width=\textwidth]{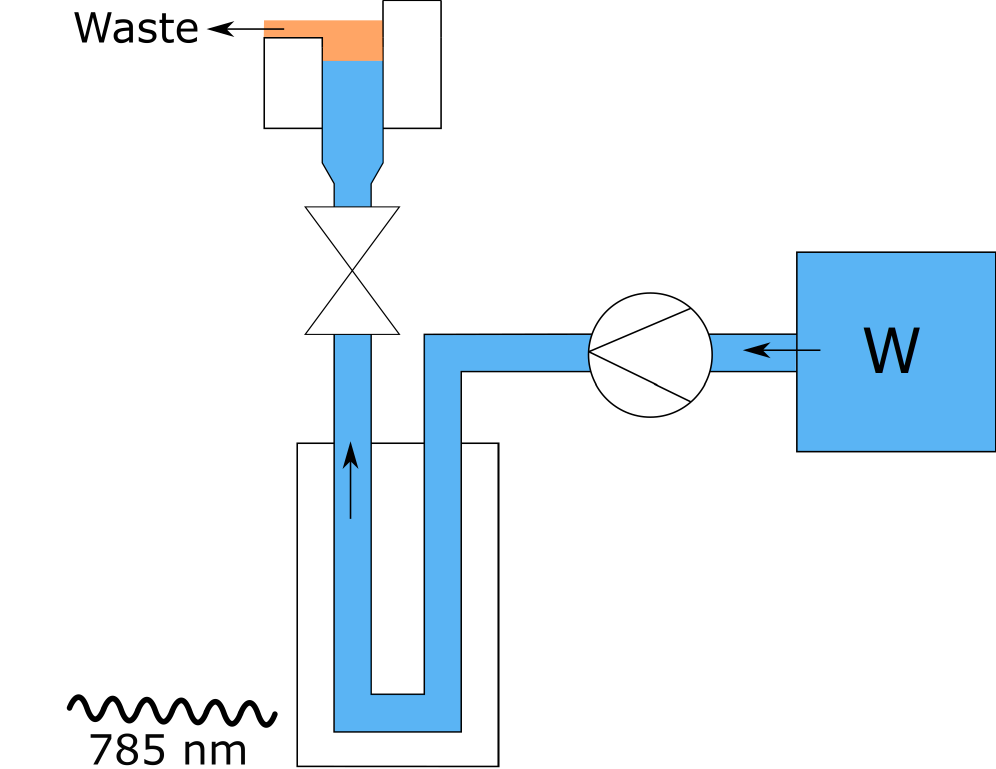}
    \caption{Cleaning for the channel}
    \label{fig:cleaning}
  \end{subfigure}

 \caption{\textit{Overview of System Components}: In Figure \ref{fig:overview} we show physical and digital components. The important stages of the sample flow during a measurement procedure are highlighted by the olive line and are illustrated in Figure \ref{fig:sample_waiting}, \ref{fig:sample_measurement} and \ref{fig:cleaning} in more detail. They involve the sampling interface (upper left corner, Figure \ref{fig:sample_waiting}, \ref{fig:sample_measurement} and \ref{fig:cleaning}), the orange sample (S), the cleaning solution (blue), the multiplexer valve (X; Figure \ref{fig:Setup}b) item $9$ and $10$), the cuvette (box at the bottom, Figure \ref{fig:Setup}b) item 12), the pump (Figure \ref{fig:Setup} item $1$ and $2$), ">" or "<" depending on pumping direction, and the reservoir of cleaning solution (W).} 
\label{fig:component_interactions}
\end{figure*}

Most of the devices discussed in Section \ref{devices_section}, whose functionalities undergo changes throughout the measurement cycle, are designated with specific services such as the spectrometer, pump and multiplexer valve. These services enable remote management of these devices. In addition, we established two more services that interact with a relational database and to predict substrate concentrations using a machine learning model.
The coordination of all these components
\begin{itemize}
    \item Spectrometer Service: Measures spectra
    \item Pump Service: Pumps back and forth 
    \item Multiplexer Service: Switches valve position
    \item Database Service: Stores and loads data
    \item Estimation Service: Infers quantities from spectra
\end{itemize}
is handled by another service, the Orchestrator, like we show in figure \ref{fig:overview}. 
Thus, each component is implemented as an independent microservice, with inter-service 
communication done via gRPC, which provides fast communication and a platform-independent API. This modular architecture simplifies modifications to any component within the system configuration. The functionalities of each service are encapsulated within its own Python package. To enhance reusability, maintenance, and dependency management, each package is equipped with a Continuous Integration pipeline for testing the code, pre-built deployment wheels, and comprehensive documentation. Further details of each package are presented in Table \ref{tab:software}.

\FloatBarrier

You can find more information on the individual services in the appendix \ref{sec:additional_software}. 
Now we will take a more detailed look at the most important items the Database-Service and the Orchestrator.

\begin{table*}
\caption{\textit{Software Components}: Here we specify the characteristics of all Python packages involved in a Raman measurement. We specify the release versions used, the fraction of code that is tested via unit tests, whether the packages use async concurrency, which operating system they can run on and which kind of API is available to reach a running server from the client side.}
\label{tab:software}
\centering
\begin{tabular}{lrrrrrr}
\toprule
Components   & \makecell[r]{Spectrometer \\ Service} & \makecell[r]{Pump \\ Service} & \makecell[r]{Multiplexer \\ Service} & \makecell[r]{Database \\ Service} & \makecell[r]{Estimation \\ Service} & Orchestrator \\
\midrule
\makecell{release} & 0.1.9 & 0.1.6 & 0.2.3 & 0.5.3 & 0.1.5 & 0.4.8 \\
\makecell[r]{test \\ coverage} & 93\% & 38\% & 44\% & 95\% & 87\%  & 90\% \\
\makecell[r]{Async} & No & No & Yes & No & Yes & Yes \\
\makecell[r]{Operating \\ System} & Windows & \makecell[r]{Linux, \\ Windows} & Windows & \makecell[r]{Linux, \\ Windows} & \makecell[r]{Linux, \\ Windows} & \makecell[r]{Linux, \\ Windows} \\
\makecell[r]{Client \\ API} & Python & \makecell[r]{CLI, \\ Python} & \makecell[r]{CLI, \\ Python} & \makecell[r]{CLI, \\ Python} & Python & \makecell[r]{CLI, \\ Python} \\
\bottomrule
\end{tabular}
\end{table*}

\subsubsection{Database Service}\label{sec:db_interaction_service_section}
The database interaction service offers a well-defined interface to the relational database, which stores all critical biolab data, especially for high-throughput experiments \cite{kaspersetz_automated_2022}. The API includes endpoints for experiment management, spectra storage and export, sample creation and annotation, metadata addition, as well as maintaining and loading machine learning models.
\\
Utilizing this service confers several benefits relative to alternative services that directly access the database on the fly.
\begin{itemize}
    \item Abstraction: the micro-service encapsulates all complex logic inherent in the database scheme which eases its usage
    \item Data Validation: the micro-service validates input data before database entry
    \item Consistency: Data is uniformly formatted across users
    \item Security: Restrict direct database access to prevent data corruption
    \item Maintainability: Changing Database logic requires updates only in one location    
\end{itemize}
More details of this service are available at \url{https://bvt-htbd.gitlab-pages.tu-berlin.de/kiwi/tf3/raman-hive/}.

\subsubsection{Orchestrator} \label{sec:orchestrator}
The orchestrator operates on a server-client model with the server controlling all services described previously. The client side receives orders from the liquid handling station or users that trigger Raman measurements. The main features of the orchestrator are to measure Raman spectra and to clean the whole setup and prepare it for a new measurement. This design simplifies control over services. More details about the orchestator service are documented at \url{https://bvt-htbd.gitlab-pages.tu-berlin.de/kiwi/tf3/raman-orchestrator/}.

\subsection{Measurement Procedure}

Both physical elements as well as software components are connected to each other in a variety of ways for liquid and data transfer, as shown in figure \ref{fig:component_interactions}. To use all components for the measurement of Raman spectra of dimension $D$, we do the steps depicted in algorithm \ref{alg:sample_measurement}. This measurement process necessitates two distinct categories of variables: Firstly, parameters that remain invariant over extended periods to ensure measurement consistency; secondly, input variables such as the number of spectra to be measured, which are subject to change between measurements. The following description outlines the procedure for an individual sample, supplemented by further graphical representations of the physical processes that occur, as illustrated in Figure \ref{fig:component_interactions}.

The measurement protocol is initiated through the orchestrator (Section \ref{sec:orchestrator}) for the $K$ sample ($1 \leq K \leq 8$), after pipetting of these $K$ samples into the interface as shown in Figure \ref{fig:sample_waiting}. Subsequently, the multiplexer (Section \ref{sec:mux_service}) is adjusted to the position $k$ to connect the sample well to the cuvette. The pump (Section \ref{sec:pump_service}) then moves the sample into the cuvette as illustrated in Figure \ref{fig:sample_measurement}. The subsequent phase of the process depends on the number of spectra $N$ to be recorded for each sample. During spectra acquisition, the sample is moved by a volume of $V_m$, which we chose to be \qty{20}{\micro \litre}. This practice helps mitigate the presence of potential air bubbles in the spectrum and reduces heat transfer. Assuming a recording time of $\tau$ seconds per spectrum, we move the sample at a flow rate $\frac{V_m}{N \tau}$, ensuring its transit until the end of recording the final spectrum. After each of the $N$ spectra is recorded (Section \ref{sec:spectrometer_service}),  they are stored in the database (Section \ref{sec:db_interaction_service_section})  alongside the relevant metadata. Subsequently, all collected spectra are sent to the estimation service (Section \ref{sec:estimation_service}), and the orchestrator sends the predictions to the database service, such that the predictions are written to the database to allow other models to use them for adjusting feeding strategies \cite{krausch_highthroughput_2022}.

Finally, in the cleaning process (see Figure \ref{fig:cleaning}), the initial action involves reversing the direction of the pump for the cleaning solution. During pumping of the cleaning solution through the flow-through cuvette, the tubes and sample are well absorbed into the overflow of the interface, which is connected to a waste water tank. Upon completion of the cleaning phase, the solution is pumped forward once more by the pull-back-volume $V_{pb}$. This volume corresponds to the amount necessary to evacuate the well of the interface, as depicted in \ref{fig:sample_waiting}, utilizing the parameter "pb". This concluding procedure ensures that the well is empty and thus prepared to accommodate the subsequent sample.
\SetKwInput{KwData}{Parameter}
\begin{algorithm}
\SetKwBlock{DoParallel}{do in parallel}{end}
\KwData{clean flow rate $Q_{c}$;
    clean volume $V_{c}$;
    sample-to-cuvette flow rate $Q_{s}$;
    sample-to-cuvette volume $V_{s}$;
    pull-back flow rate $Q_{pb}$;
    pull-back volume $V_{pb}$;
    move-sample volume $V_{m}$
}
\KwIn{number of samples $K$;
    measurement time $\tau$;
    number of scans $N$;
    substrates to predict $S$;
    model id $M$;

}
\KwResult{Spectra $X \in \mathbb{R}^{K \times N \times D}$ and Estimations $\hat{Y}(X) \in \mathbb{R}^{K \times S}$ }
\For{$k\gets1$ \KwTo $K$ }{
    set multiplexer to $k$-th position ; \\
    pump forward $V_{s}$ at flow rate $Q_{s}$;\\
    \DoParallel{
        pump forward $V_{m}$ at flow rate $\frac{V_{m}}{N \tau}$;\\
        \For{$n\gets1$ \KwTo $N$ }{
            record spectrum $X_{nk:}$ for $\tau$ seconds; \\
            store spectrum $X_{nk:}$ in the database;\\
        }
        predict $\left(\hat{Y}_s(X_{nk:})\right)_{s=1, \dots S} $; \\
        store predictions $\hat{Y}$ in the database; \\

    }
    pump backward  $V_{c}$ at flow rate $Q_{c}$;\\
    pump forward $V_{pb}$ at flow rate $Q_{pb}$;\\
}
\caption{\textit{Measuring $K$ Samples}: We describe the steps during the measurement of Raman spectra of $K$ samples.}
\label{alg:sample_measurement}
\end{algorithm}

\subsection{Calibration Procedure}  \label{calibration_procedure}

In order to ascertain the optimal configuration of sample-to-cuvette volume, move-sample volume, pull-back volume, and sample volume, both the mean intensity of Raman spectra and their corresponding overall standard deviation are systematically assessed. Therefore, we want to maximize the spectral intensity and minimize the standard deviation to ensure reproducibility. We chose solutions of \qty{80}{\gram \per \litre} Glucose (D-(+)-glucose monohydrat, Carl Roth, Karlsruhe, Germany) and \qty{50}{\gram \per \litre} MgSO$_4$ (Magnesium Sulfate Heptahydrate, Carl Roth, Karlsruhe, Germany) as calibration substances due to their high Raman activity and distinct peaks. The calibration process begins with manual testing to set initial parameters, followed by a systematic parameter search of the parameter grid using the Tecan Liquid Handling Station (LiHa). The mean intensity for glucose is measured over the wavenumber range of \qtyrange{1000}{1500}{\per \cm} for its characteristic double peaks, and for MgSO$_4$, the range is \qtyrange{960}{1000}{\per \cm}, as illustrated in Figure \ref{fig:raw_spectra}.

We need to specify the parameters of the algorithm \ref{alg:sample_measurement} for optimal spectra quality. Our setup involves 7 parameters with numerous potential values, making it impractical to test all combinations. Therefore, we focus on sample-to-cuvette and pull-back volume, critical for accurate liquid placement in the cuvette. As these crucial parameters are interdependent, we chose a data-driven approach. 

But first we looked for decent settings for the other $5$ parameters individually. For flow-rates $Q_c$, $Q_s$, and $Q_{pb}$, we opted for the highest possible values until issues such as spillover or reduced reproducibility occurred. These high flow-rates enabled us to reduce the overall meaurement time per sample. We set the move-sample volume at \qty{20}{\micro \litre} and ensured consistent spectra during multiple recordings while pumping $V_m$. For clean volume, we chose \qty{1.25}{\milli \litre} after testing with a highly concentrated solution of magnesium sulfate in well one and water in well 2, ensuring that we do not observe a residual sulfate peak at \qty{980}{\per \centi \meter} in the second well post-measurement. A sample volume of \qty{50}{\micro \litre} was used to adequately fill the cuvette which can take up to  \qty{18}{\micro \litre} and accommodate the move-sample volume \qty{20}{\micro \litre}.\\
Subsequently, we used the values mentioned above to systematically test the sample-to-cuvette and pull-back volume. The range for the sample-to-cuvette volumes was \qtyrange{75}{140}{\micro \liter} in increments of \qty{5}{\micro \litre}. Simultaneously, we tested the pull-back volumes over a range of \qtyrange{350}{400}{\micro \litre}, also in \qty{5}{\micro \litre} increments. For each combination of volumes we measured at least $40$ spectra. Testing both of these parameters is crucial as they are interdependent. \\

\subsection{Maintenance}

To uphold the quality of the measurements, regular maintenance is imperative. Of particular importance is the cleaning of the system, which involves flushing the setup with water and eliminating residues using anti-static cotton swabs saturated with isopropanol. The integrity of the spectral data can be verified by analyzing samples of demineralized water and comparing the configuration and intensity of the obtained spectra with a high-quality reference spectrum. 
When working with cells, afterwards the system should be flushed with a 5\% hydrochloric acid solution to eliminate mineral deposits, metal oxides, and other contaminants. Subsequently, we performed repeated cleaning of the apparatus with demineralized water.

\section{Results}

We show the best measurement conditions that we obtained from our calibration procedure, get the durations of the individual components, and show how we used them to measure spectra during a fermentation of \textit{Escherichia coli}.

\subsection{Calibration Results}
\begin{figure*}[h!]
    \centering
        \begin{subfigure}{\textwidth}
            \centering
            \includegraphics[width=\textwidth]{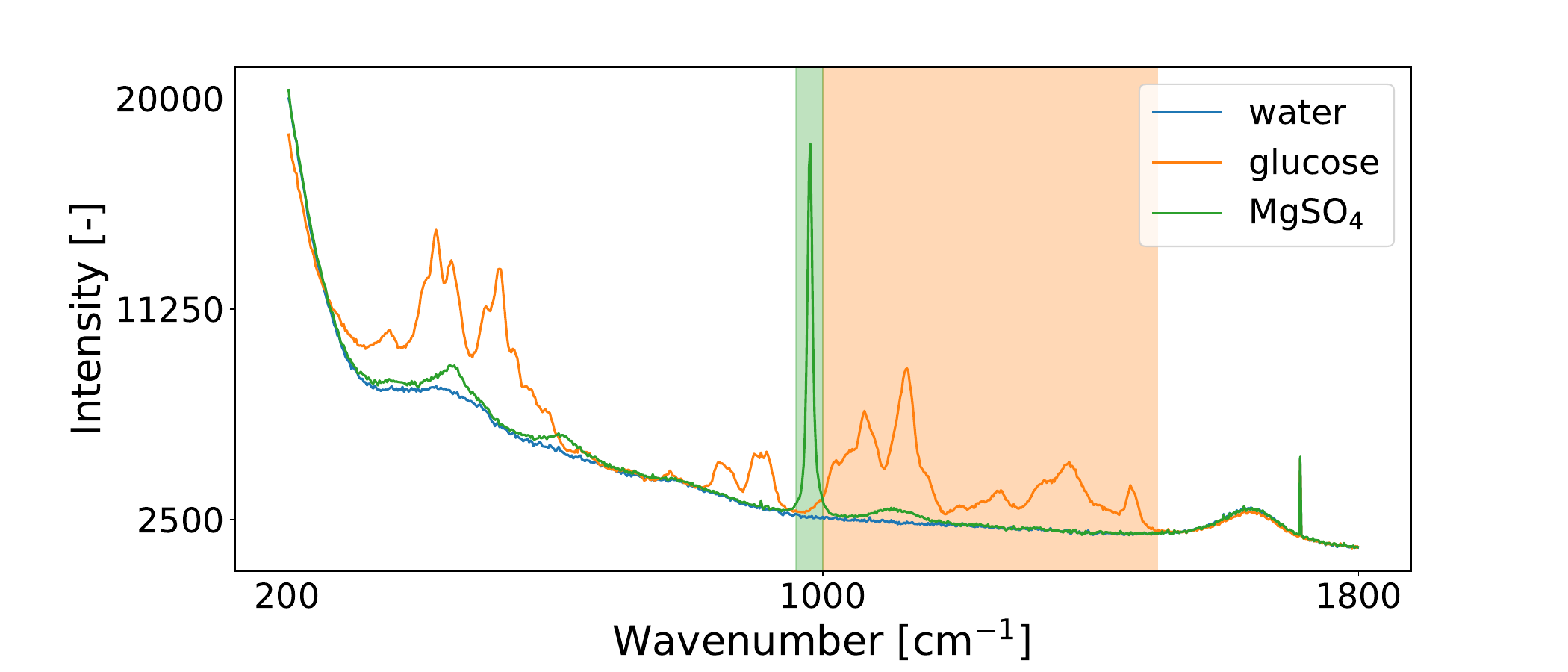}
            \caption{Raw spectra of water, glucose, and MgSO$_4$.}
            \label{fig:raw_spectra}
        \end{subfigure}
        \hfill
        \begin{subfigure}{0.49\textwidth}
            \includegraphics[width=\textwidth]{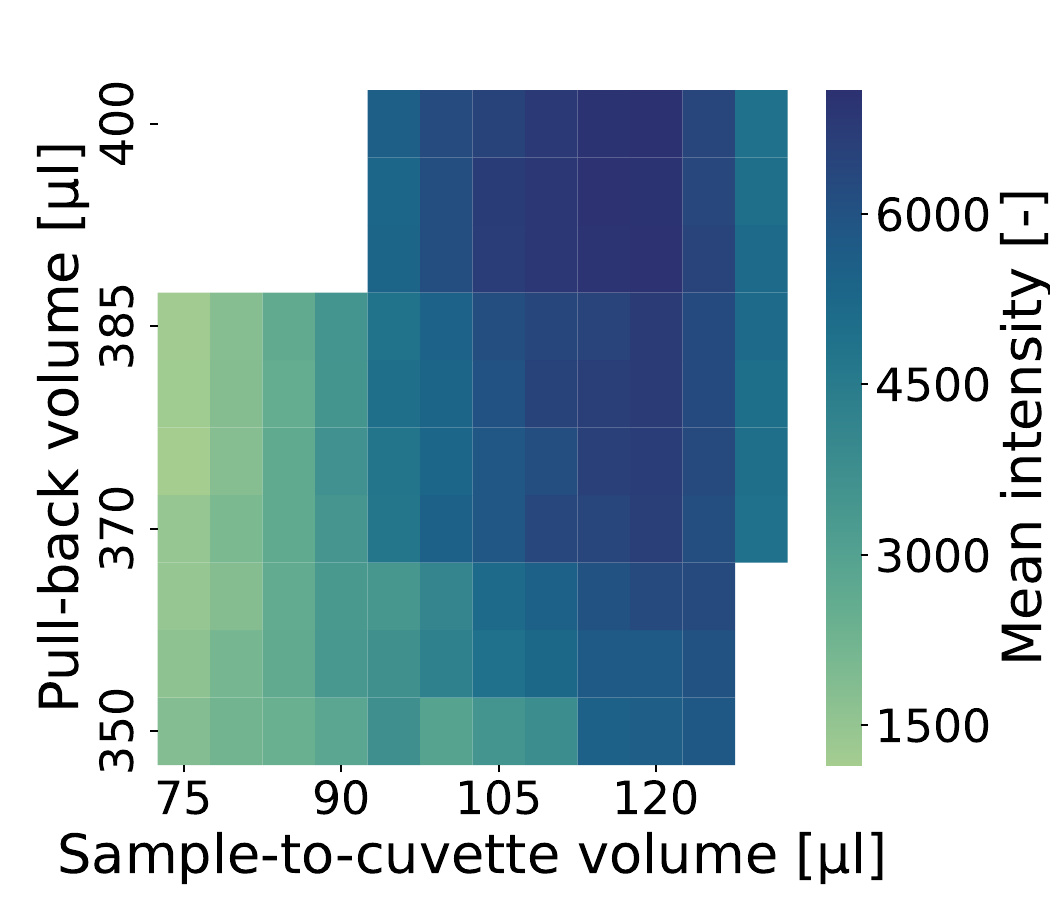}
            \caption{Mean intensity heatmap for MgSO$_4$ (960--1000 cm$^{-1}$).}
            \label{fig:mgso4_heatmap}
        \end{subfigure}
        \hfill
        \begin{subfigure}{0.49\textwidth}
            \includegraphics[width=\textwidth]{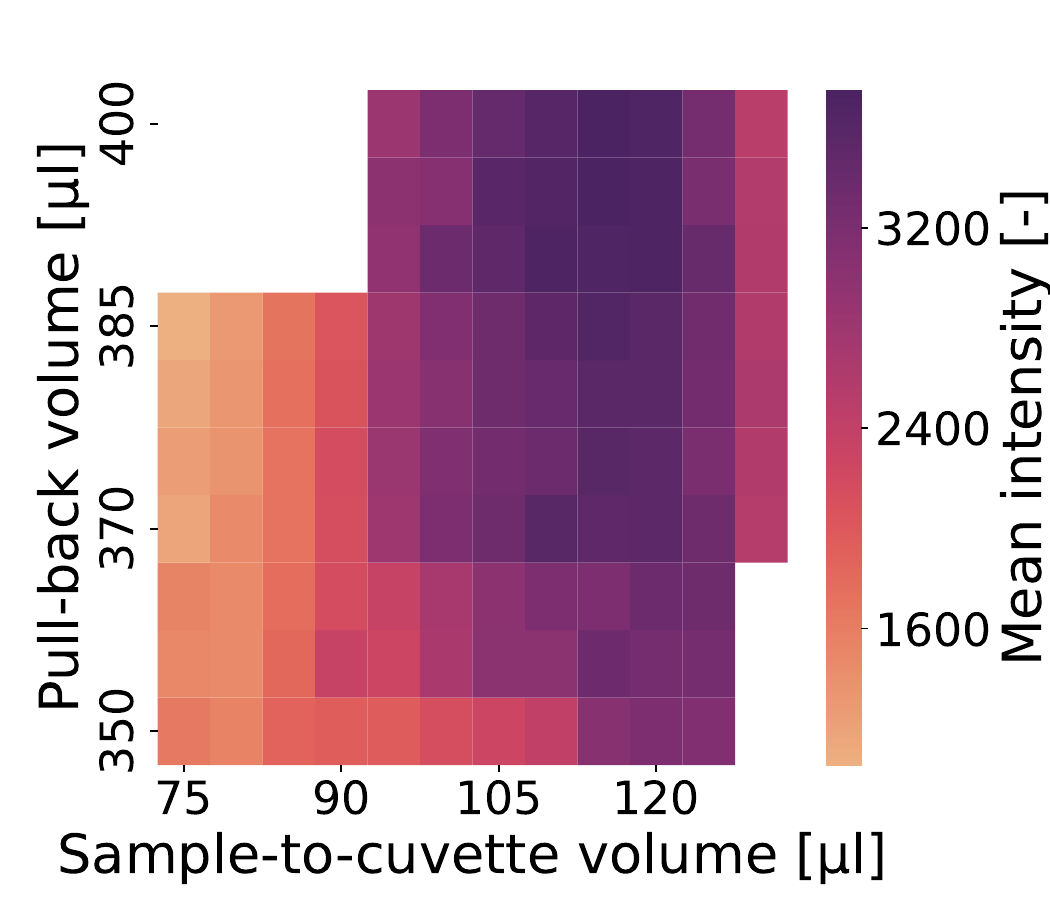}
            \caption{Mean intensity heatmap for glucose (1000--1500 cm$^{-1}$).}
            \label{fig:glucose_heatmap}
        \end{subfigure}
    \caption{Raw spectra and corresponding mean intensity heatmaps for glucose and MgSO$_4$.}
    \label{fig:combined}
\end{figure*}

This section examines the results of the calibration method that aims to provide reliable high-quality spectra from the analytes. According to the calibration procedure explained in Section \ref{calibration_procedure} we depict the signal intensity in
Figure \ref{fig:combined} under various conditions. These measurements are conducted for both glucose (Figure \ref{fig:glucose_heatmap}) and magnesium sulfate (Figure \ref{fig:mgso4_heatmap}) to determine if the findings apply to various substances. \\
For both substances, the spectral intensity increases with greater pullback volumes. This trend occurs because larger pullback volumes decrease the amount of wash solution remaining in the wells after the washing steps, resulting in fewer dilutions of the analyte.
For the sample-to-cuvette volume the right amount makes sure that the sample is placed directly in front of the laser (Figure \ref{fig:sample_measurement}).\\
We observe a high signal intensity for the sample-to-cuvette volume to be \qtyrange{105}{120}{\micro \litre}. Volumes below this range have a lower intensity, probably resulting from an insufficient amount of sample reaching the cuvette. On the other hand, volumes above \qty{120}{\micro \litre} exhibit a steep decline in signal intensity probably caused by the increase in the fraction of the sample that is pumped beyond the cuvette.\\
For glucose in Figure \ref{fig:glucose_heatmap} we observe the highest signal intensity at \qty{115}{\micro \litre} sample-to-cuvette and \qty{395}{\micro \litre} pull-back volume. In contrast, in Figure \ref{fig:mgso4_heatmap} magnesium sulfate shows the best calibration at a pull-back volume of \qty{400}{\micro \litre} and a sample-to-cuvette volume of \qty{120}{\micro \litre}. However, the setting with a \qty{115}{\micro \litre} sample-to-cuvette volume and a \qty{395}{\micro \litre} pull-back volume offers the third highest intensity. 
As a result, we opted for this setting to align with the glucose calibration.\\
Moreover, we took the standard deviation of the signal intensities into account (Figure \ref{fig:cal_stds}. We observe that around the setting with a \qty{115}{\micro \litre} sample-to-cuvette volume and a \qty{395}{\micro \litre} pull-back volume is generally low. This indicates a high reliability of our measurement. \\

\subsection{Measurement Duration}
In this section, we outline the time required for each step of the Raman measurement process that we described in Algorithm \ref{alg:sample_measurement} in more detail.
\begin{itemize}
    \item Duration switching valve position \qty{1.3}{\second}
    \item Duration pumping sample to cuvette \qty{3.6}{\second}
    \item duration cleaning channel \qty{10}{\second}
    \item duration pull back \qty{4.74}{\second}
    \item time spent in software including prediction \qty{0.56}{\second}
    \item total overhead time \qty{20.2}{\second}
    \item measuring a Raman Spectrum \qty{10}{\second}
\end{itemize}
As all the steps except the measurement duration per spectrum have a fixed duration, i.e. we always have an overhead around \qty{20.2}{\second} that comes with each sample. The exposure time of each Raman spectrum and the number of spectra to be measured are free to choose for the user. In our experiments we used \qty{10}{\second}  per spectrum and recorded two spectra per sample, but one can increase both numbers to improve the signal to noise ratio.

\subsection{Measurement during a Fermenation} \label{sec:spectra_during_fermenation}
\begin{figure*}[h]
\centering
\includegraphics[scale=0.48]{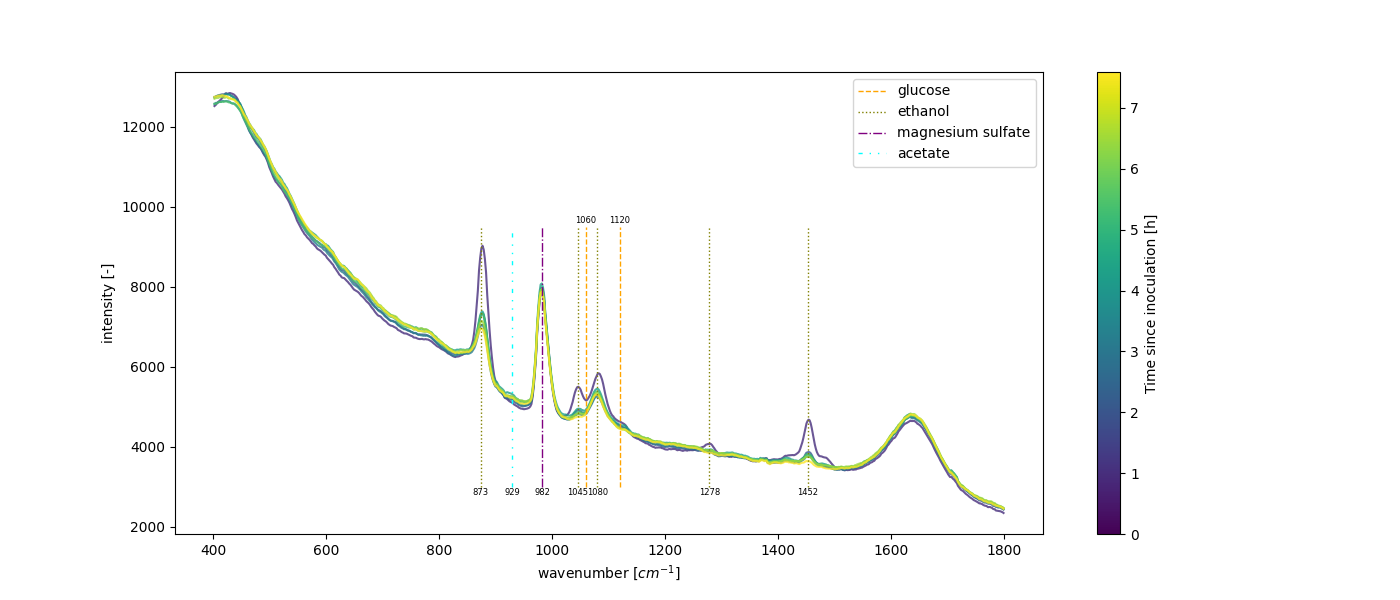}
\caption{\textit{Raman Spectra during E. coli Fermentation:} Here we see the Raman spectra of the supernatent  of one E. Coli fermentation measured with the setup described in this paper. The numbers above or below the vertical lines indicate the peak positions obtained from the literature \cite{emin_raman_2020,wang_sulfates_2006,frost_raman_2000,mathlouthi_laser-raman_1980}}
\label{fig:experiment_spectra}
\end{figure*}
We used the setup described herein in a fermentation process of \textit{Escherichia coli}, encompassing a batch of approximately \qty{4}{\hour} and a glucose-limited feed batch of \qty{3}{\hour}. The experiment was carried out in a manner consistent with the procedure described in \cite{krausch_highthroughput_2022}. Due to the high cell density causing turbid samples and resulting in low Raman signal intensity, we centrifuge the samples prior to transferring the supernatant into the sampling interface. \\
In Figure \ref{fig:experiment_spectra} we observe a stable baseline in the spectra, along with temporal variations in peak intensities. In particular, several distinguished peaks can be linked to ethanol \cite{emin_raman_2020}, which serves to disinfect the needles of the liquid handling station. Over time, the intensity of these ethanol peaks decreases, corresponding to an increase in the cell count that can consume some of the ethanol during the sampling procedure. Observations of the acetate peak \cite{frost_raman_2000} indicate a slight increase in acetate levels, an aerobic overflow metabolite and anaerobic by-product, over the course of the fermentation. Glucose, being the primary carbon source, diminishes over time. Magnesium sulfate, which acts as a cofactor to enhance microbial substrate-to-product conversion \cite{gotsmy_sulfate_2023}, shows a slight decrease.

\section{Discussion}

In this study, we developed and tested an automated procedure for measuring Raman spectra at-line with a liquid handling robot fully automated during high-throughput fermentations, calibrations procedures or offline analytics.

\subsection{Potential Use Cases}

We developed the setup for the usage within automated fermentations for high-throughput bioprocess development of up to $48$ mini bioreactors in parallel \cite{}. Therefore, everything is optimized regarding automation, reliability and speed. However, one can use the system for monitoring enzymatic reactions as well as within biocatalysis or organic chemistry. 

Raman spectroscopy can provide real-time, label-free detection of molecular changes by analyzing vibrational fingerprints of substrates, intermediates and products, and can thus greatly help in understanding how enzymes act on the molecular basis \cite{carey_advances_2021}. 
For example, Raman spectroscopy can track redox changes in coenzymes such as NADH/NAD during dehydrogenase-catalyzed reactions \cite{chen_classical_1987} or detect structural modifications in peptides \cite{sahoo_analysis_2016} and polysaccharides \cite{he_principal_2020} during enzymatic hydrolysis. It can also be used to analyze the concentrations of reaction components. 
Additionally, this setup supports advanced Raman techniques like Surface Enhanced Raman Spectroscopy (SERS), which necessitates adding substances such as silver nanoparticles to the sample to amplify the signal \cite{lin_raman_2021}. SERS increases the sensitivity and resolution of Raman spectroscopy and thus allows for deeper explorations of the sub-microscopic domain \cite{xia_deciphering_2024}. 

In all these applications, our Raman system, integrated into the liquid handling station, facilitates diverse pretreatment methods for samples. For example, one could dilute a sample to a specific optical density for characterizing inclusion bodies or homogenize samples to analyze intracellular contents. Alternatively, centrifuging the samples allows separation of cells from the supernatant, enabling the analysis of media components. 

\subsection{Implications for Usage of Machine Learning Models}

The fully automated setup presented here for measuring Raman spectra streamlines the entire measurement process, thereby facilitating the acquisition of larger datasets. The availability of more data will promote the development of more complex machine learning models such as \cite{arend_detection_2020}. Currently, complex models such as neural networks are primarily addressing classification problems \cite{luo_deep_2022} in the domains of medicine, biology, and biotechnology \cite{sohn_singlelayer_2020, wu_deep_2021, hu_raman_2022, dong_practical_2019, maruthamuthu_raman_2020} that require fewer data compared to regression problems. With bigger datasets being available, we will be able to combine them to train more sophisticated neural network architectures that, for instance, tackle the regression problem of inferring concentrations from Raman spectra in complex biotechnological applications. For example, convolutional neural network (CNN) architectures such as ResNet \cite{khan_survey_2020, bitra_machine_2023} have gained popularity in the analysis of Raman spectra \cite{ho_rapid_2019, chen_serum_2020} or transformer architectures that are more common in image or spectral data \cite{vaswani_attention_2023,chang_rat_2024,koyun_ramanformer_2024,he_masked_2021}. Thus, uniform measurement conditions in the cuvette (Section \ref{cuvette_device}) for all types of use cases promote data efficiency. It harmonizes the Raman spectra patterns across various applications, which reduces the necessity for comprehensive calibration data in new scenarios, thereby working towards establishing a foundation model for Raman spectra.

\subsection{Design Considerations}

With regard to design considerations, there are alternative approaches that could be employed, necessitating an explanation of our thought process. For example, we have chosen to keep the tubes filled with cleaning solution. This decision is grounded in the fact that the pump achieves more consistent results when pumping liquid rather than a combination of samples and air, particularly when the majority of the tube is occupied by air. Consequently, the placement of the sample in the cuvette is more consistent with the presence of liquid in the tubes, thereby resulting in more reliable spectra. 
With respect to the measurement procedure, we considered various alternatives. In particular, the sequence of the distinct steps is flexible; it is not imperative to perform the cleaning of each channel immediately following the measurement of the sample. An alternative approach involves measuring the samples across all channels prior to conducting the cleaning for each channel individually in a stop flow manner. That option would decrease the lag time between the samples placed in the interface and the Raman spectra of the last sample being measured. Nevertheless, there are two significant drawbacks to this method. Firstly, the cleaning of the cuvette and the tubes between the cuvette and the multiplexer valve is constrained, as it relies solely on the washing solution already present in the tubes of the succeeding channel between the interface and the cuvette. Therefore, depending on the concentrations of substances present in the samples the amount of washing solution may not suffice. Secondly, this strategy necessitates altering the position of the multiplexer $2 K$ times per measurement cycle, in contrast to $K$ times required in the current arrangement, thereby resulting in a saving of \qty{1.3}{\second} per channel, which is the average time the multiplexer valve needs to change its position.

\section*{Author Contributions}
According to \href{https://credit.niso.org/}{CRediT} for the descriptions of standardized contributions, we have the following distribution. Christoph Lange: Conceptualization, Data Curation, Formal Analysis, Investigation, Methodology, Software, Visualization and Writing - Original Draft; Simon Seidel: Conceptualization, Methodology, Resources, Visualization and Writing - Original Draft; Madeline Altmann: Conceptualization, Data Curation, Formal Analysis, Investigation, Methodology, Visualization and Writing - Original Draft; Daniel Stors: Conceptualization, Data Curation, Formal Analysis, Investigation, Methodology, Visualization and Writing - Original Draft; Annina Kemmer: Investigation, Resources, Supervision and Writing - Review \& Editing; Linda Cai: Investigation, Resources, Supervision and Writing - Review \& Editing; Stefan Born: Conceptualization, Supervision and Writing - Review \& Editing; Peter Neubauer: Supervision, Funding Acquisition and Writing - Review \& Editing; M. Nicolas Cruz Bournazou: Supervision, Funding Acquisition and Writing - Review \& Editing;

\section*{Acknowledgements}
CL, SiS, MA, DS, SB, PN, and NK gratefully acknowledge the financial support of the German Federal Ministry of Education and Research (01DD20002A - KIWI biolab). AK received financial support from Berliner ChancengleichheitsProgramm (BCP) as part of the graduate program DiGiTal. LC received financial support from GlaxoSmithKline Biologicals SA as part of a research collaboration agreement.
The authors acknowledge support by the Open Access Publication Fund of TU Berlin.

\section*{Conflict of interest}
The authors declare no conflicts of interest.

\bibliographystyle{rss}
\bibliography{integration}

\newpage
\appendix

\section{Additional Software Components}\label{sec:additional_software}
\subsection{Spectrometer Service}\label{sec:spectrometer_service}
The spectrometer service is specifically tailored for the Metrohm i-Raman Plus spectrometer as described in Section \ref{spectrometer_device}. It offers an API that allows measurements of spectra by specifying either the laser intensity or the measurement duration. More information on this service can be found at \url{https://bvt-htbd.gitlab-pages.tu-berlin.de/kiwi/tf3/raman-spectrometer-server/}.

\subsection{Pump Service}\label{sec:pump_service}
The pump service facilitates communication with the specified Ismatec pump (Section \ref{pump_device}) with which communication occurs through a serial connection. This service provides an API to the pump, allowing the configuration of parameters such as flow rate, volume, direction, delay, and the option to wait for pump completion. In addition, it offers an interface to modify the default settings and perform a health check. Additional details on this service can be found in \url{https://bvt-htbd.gitlab-pages.tu-berlin.de/kiwi/tf3/ismatec-pump/}.

\subsection{Multiplexer Service}\label{sec:mux_service}
The multiplexer valve service facilitates altering the valve's position. It is tailored for the Elveflow valve, which is elaborated upon in Section \ref{mux_device}. The API for adjusting the valve includes a feature to set the direction, as well as a health check functionality. Additional information about this service can be found at \url{https://bvt-htbd.gitlab-pages.tu-berlin.de/kiwi/tf3/multiplexer-valve/}.

\subsection{Estimation Service} \label{sec:estimation_service}

The estimation service infers substance concentrations from a Raman spectrum. At startup, it loads specified models from the database via the API in Section \ref{sec:db_interaction_service_section}, performs preprocessing, and allows predictions for all trained parameters. More details are available at \url{https://bvt-htbd.gitlab-pages.tu-berlin.de/kiwi/tf3/raman-estimate/}.

Here we add the standard operating procedure that is used in our lab when using the system. 

\section{Standard Deviation of the Calibration Results}

\begin{figure*}[h] 
        \centering
        \hfill
        \begin{subfigure}[a]{0.49\textwidth}
        \includegraphics[width=\textwidth]{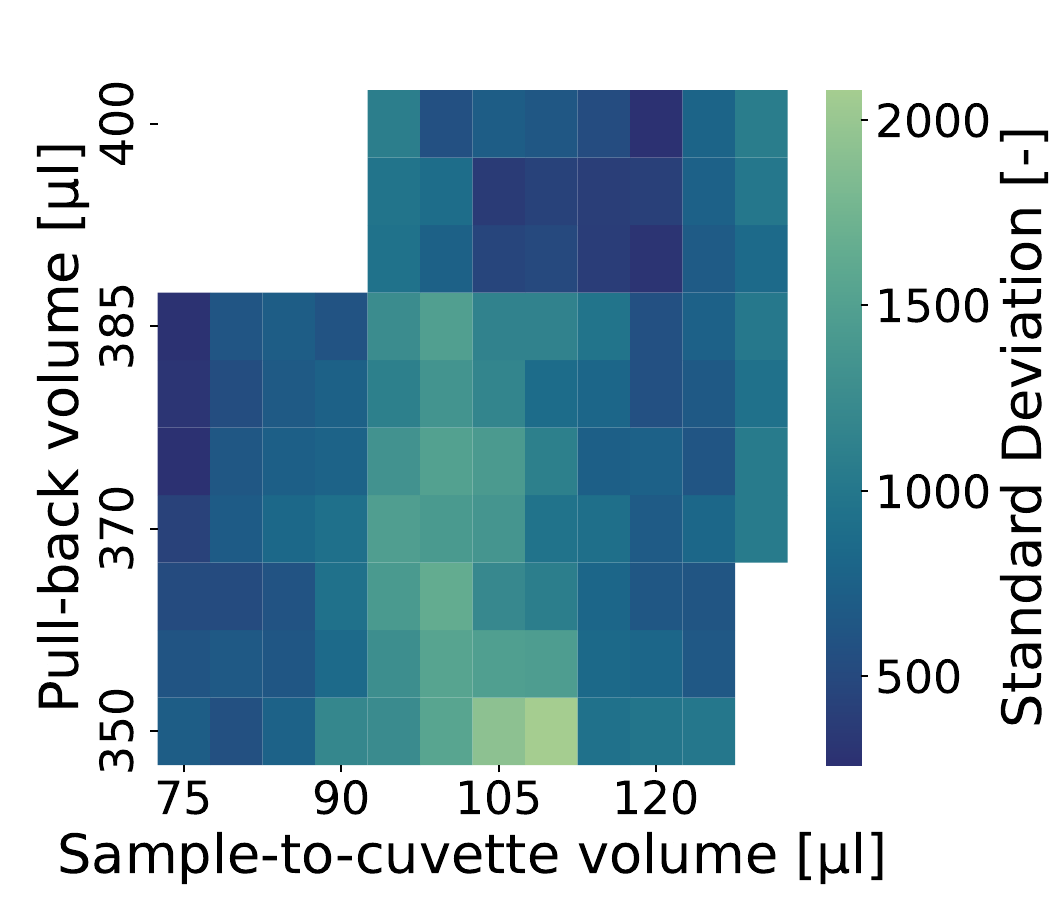}
        \label{fig:mg_cali}
        \end{subfigure}
        \begin{subfigure}[a]{0.49\textwidth}
        \includegraphics[width=\textwidth]{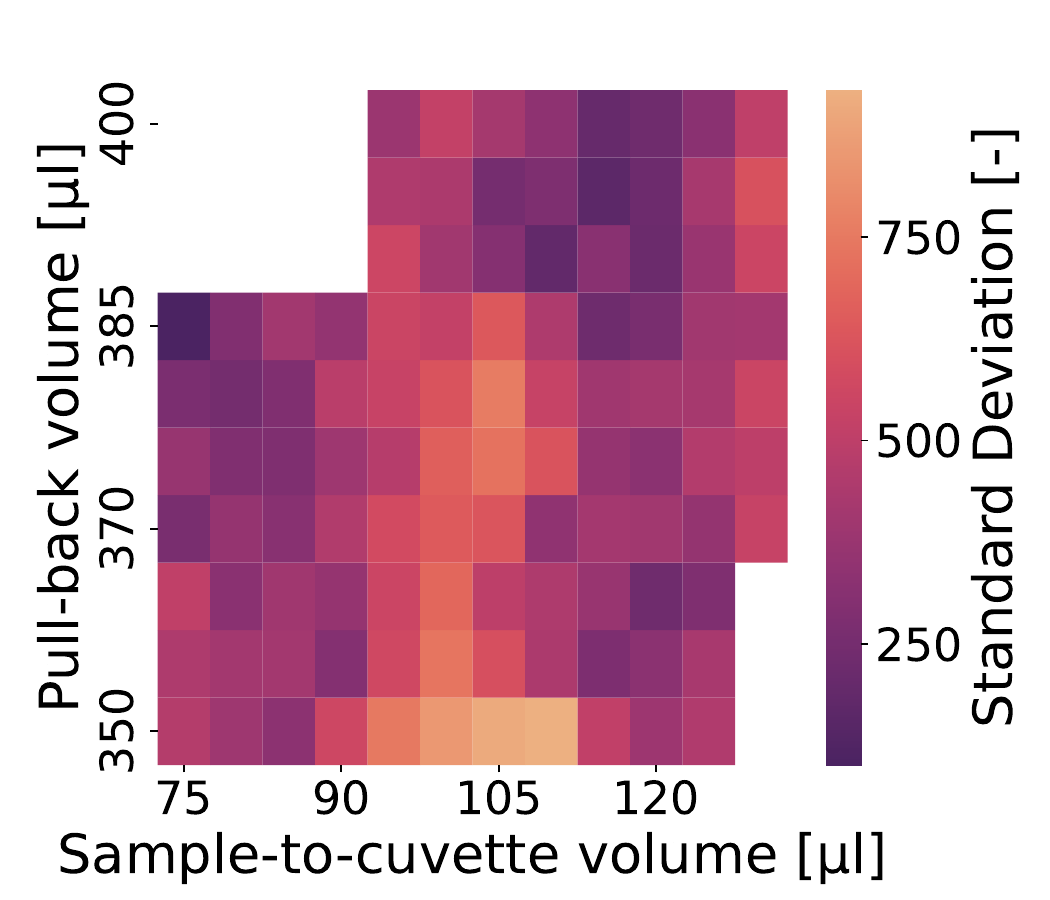}
        \label{fig:glucose_cali}
        \end{subfigure}
        \caption[Standard Deviations of the Calibration of the Raman Interface.]{\textit{Standard deviation of Raman calibration} For magnesium sulfate (a) and glucose (b)  we measured at least $40$ samples using different settings for the pull-back volume and the sample-to-cuvette volume. Here we show the standard deviation of the signal.}
        \label{fig:cal_stds}
\end{figure*}

Here we provide the standard deviation of the signal intensites during the calibration runs. In Figure \ref{fig:cal_stds} we show the results for each setting for glucose and magnesium sulfate.  

\section{Standard Operating Procedure}

This is an SOP for the use of the Raman measurement setup during a cultivation in the 2mag system. It gives instructions on how to prepare the system for an experiment and how to handle it during an experiment. It also gives short maintenance instructions and a short troubleshooting help.

\subsection{Raman Paramters}\label{parameters}
Table \ref{tab:parameters} gives an overview of the experiment and how the different variables of the system have been set. The standard values for most of the variables are given and should not need to be changed for a standard experiment.
\begin{table*}[h]
\caption{Raman parameters (standard values given)}
\label{tab:parameters}
\begin{tabular}{>{\raggedright\arraybackslash}p{0.3\textwidth}>{\raggedright\arraybackslash}p{0.3\textwidth}>{\raggedright\arraybackslash}p{0.3\textwidth}}
\hline
\textbf{Parameter} & \textbf{Value} & \textbf{Details / Comments} \\ 
\hline
\textbf{Model (device id)} & &  Refer to SQL data base to find device-id of the model that estimates concentrations from spectra. \\ 
\hline
\textbf{run-id} & & Refer to SQL data base to select your created run or create new run with raman-hive commands (see \href{https://bvt-htbd.gitlab-pages.tu-berlin.de/kiwi/tf3/raman-hive/}{documentation}) \\ 
\hline
\textbf{Substances of interest} & Glucose, Acetate, Phosphate, Glycerol, Nitrate, Magnesium Sulfate & The substances that the model will predict the concentration of \\ 
\hline
\textbf{Measurement method} & manual/automatic & Manual: samples need to be pipetted into the pipetting interface by hand; automatic: samples are pipetted by the liquid handling robot \\ 
\hline
\textbf{sample-to-cuvette-pump-volume [\SI{}{\litre}]} & 1.15e-4 & Volume pumped to bring the sample from the interface to the cuvette for measuring \\ 
\hline
\textbf{pull-back-pump-volume [\SI{}{\litre}]} & 3.95e-4 & Volume to empty the well after flushing \\ 
\hline
\textbf{clean-pump-volume [\SI{}{\litre}]} & 1.25e-3 & Volume pumped for cleaning after a sample has been measured \\ 
\hline
\textbf{sample-to-cuvette-flow-rate [\SI{}{\litre \per \minute}]} & 7.5e-3 & flow-rate corresponding to sample-to-cuvette-pump-volume \\ 
\hline
\textbf{pull-back-flow-rate [\SI{}{\litre \per \minute}]} & 5e-3 & flow-rate corresponding to pull-back-pump-volume\\ 
\hline
\textbf{clean-flow-rate [\SI{}{\litre \per \minute}]} & 7.5e-3 & flow-rate corresponding to clean-pump-volume \\ 
\hline
\textbf{move-sample-volume [\SI{}{\litre}]} & 2e-5 & Volume the sample is pumped forwards during measurement \\ 
\hline

\textbf{duration} [\unit{\second}] & 10 & Duration of one scan \\ 
\hline
\textbf{num\_scans} & 2 & Number of scans performed per measurement \\ 
\hline
\textbf{sample-volume [\SI{}{\micro \litre}]} & 50 & Volume of sample pipetted into the interface \\ 
\hline
\textbf{number-of-wells} & 8 & Number of wells used for the measurement \\ 
\hline
\textbf{Time and date of water check} & &  The time and date at which the water check has been performed \\ \hline
\textbf{Run-id of the water check} & & The run-id used to store the spectra for the water check before the cultivation \\ 
\end{tabular}
\end{table*}

\subsection{Pre-experiment preparation}\label{pre-experiment}
In Table \ref{tab:pre-experiment} we present in more detail the preparations that need to be done a few days before the experiment.

\begin{table*}[h]
\caption{Pre-experiment preparation}
\label{tab:pre-experiment}
\begin{tabular}{>{\raggedright\arraybackslash}p{0.15\linewidth}>{\raggedright\arraybackslash}p{0.75\linewidth}}
\textbf{Step} & \textbf{Details}  \\ 
\hline
    Train model & Training a model and save it into the database\begin{itemize}
        \item Check out the notebook to train a model:  \url{https://git.tu-berlin.de/bvt-htbd/kiwi/tf3/raman_master/-/blob/main/notebooks/neural_networks/train_nn_that_predicts_std_mock_dataset.ipynb?ref_type=heads}
        \item Store model in database \url{https://bvt-htbd.gitlab-pages.tu-berlin.de/kiwi/tf3/raman-hive/cli.html#raman-hive-client-store-model}
        \item Remember the device-id of your model
    \end{itemize}  \\ 
\hline
Incorporate Raman script into your Tecan script & We wrote a script that reads the appropriate information from a configuration file and creates a command for conducting Raman measurements via EVOWARE software \url{https://git.tu-berlin.de/bvt-htbd/facility/tecan_write_gwl/-/blob/master/Tecan_write_gwl/Tecan_write_gwl_for_Raman_measurement.py?ref_type=heads} \\ 
\hline
Set Raman parameters & If you are using the automatic sampling method, the variables for the Raman measurements need to be set in the config file for the run (config\_Tecan\_write\_gwl.json) The parameters used should ideally be the same as the ones used to generate the data with which the model has been trained. \\ 
\hline
Deep clean of the Raman setup & see  \hyperref[maintenance]{maintenance} \\ 
\hline
Check liquid levels of canisters & refill system liquid canister with distilled water; empty waste canister \\ 
\hline
\end{tabular}
\end{table*}

\subsection{Cultivation}\label{cultivation}
In this section the steps necessary to do on the day of the cultivation are detailed. This section is split into three parts: the \hyperref[preparations]{preparations to do right before the run}, what to do \hyperref[during-run]{during the run} and the \hyperref[after-run]{shut down and cleaning procedure after the run}

\subsubsection{Preparations before a Cultivation}\label{preparations}
In Table \ref{tab:preparations} we show the steps that need to be performed immediately before the experiment in more detail.

\begin{table*}[h]
\caption{Preparations right before cultivation}
\label{tab:preparations}
\begin{tabular}{>{\raggedright\arraybackslash}p{0.15\linewidth}>{\raggedright\arraybackslash}p{0.75\linewidth}}
\textbf{Step} & \textbf{Details}\\ 
\hline
Start Devices & Switch on the devices: \begin{enumerate}
    \item Spectrometer (switches on power supply and back of device)
    \item Pump (switch on back of device)
    \item Multiplexer valve (switch on side of device)
    \end{enumerate}
Don't forget to open the shutter of the laser (switch on top of the light guide) \\ 
\hline
Start servers & In the terminal of the respective project, start the servers for the individual devices, the database and orchestration service. Start the Raman-orchestrator service last: \begin{enumerate}
    \item Raman hive
    \item Multiplexer-valve
    \item Ismatec-pump
    \item Spectrometer
    \item Raman-estimate (optional for only measuring spectra)
    \item Raman-orchestator
\end{enumerate}
The commands for starting the services and more details can be found in the \href{https://bvt-htbd.gitlab-pages.tu-berlin.de/kiwi/tf3/raman-orchestrator/start.html}{documentation}. Check if errors occur after starting a microservice. If yes, check if the command is correct and if the respective device is switched on properly. \\
\hline
Flush system with water & Flush the system with water to flush out any air bubbles trapped in the system. The command you can use may look like this: \textit{raman-client queue-washing-tasks \texttt{--}grpc-channel 127.0.0.1:50051 \texttt{--}clean-pump-volume 5e-3 \texttt{--}clean-flow-rate 0.75e-2 \texttt{--}pull-back-pump-volume 3.85e-4 \texttt{--}pull-back-flow-rate 5e-3 \texttt{--}number-of-wells 8}. Using this command once takes approx. \SI{6}{\minute}. More information can be found in the \href{https://bvt-htbd.gitlab-pages.tu-berlin.de/kiwi/tf3/raman-orchestrator/cli.html#interface}{documentation}. Flushing at least twice may gives better results. \\ 
\hline
Check Spectrum Quality  & Measure water in the system and compare the spectrum to a good quality water spectrum. Use \href{https://git.tu-berlin.de/bvt-htbd/kiwi/tf3/raman_master/-/blob/main/notebooks/water_check.ipynb?ref_type=heads}{this notebook} for easy evaluation and for more details on the process. If there are any bumps or other irregularities in the spectrum or the spectrum has a low intensity repeat the flushing step or perform a more thorough cleaning (see \hyperref[maintenance]{maintenance}). If the spectrum looks good, the system is ready for operation \\ 
\hline
\end{tabular}
\end{table*}

\subsubsection{Conducting a Run}\label{during-run}
Table \ref{tab:during-run} details the steps necessary in order to measure Raman spectra during the cultivation run. It distinguishes between 3 modes: \textit{manual}, where the samples are pipetted manually into the interface, \textit{automatic}, where the samples are pipetted by the liquid handling arm and \textit{offline Raman}, where the samples are not measured during the run, but stored until measurement at a later time point.

\begin{table*}[h]
\caption{During the run}
\label{tab:during-run}
\begin{tabular}{>{\raggedright\arraybackslash}p{0.15\linewidth}>{\raggedright\arraybackslash}p{0.75\linewidth}}
\textbf{Step} & \textbf{Details}  \\ 
\hline
Sampling & \begin{enumerate}
    \item Label the plates with: runID\textless{}runID\textgreater{}-\textless{}date\textgreater{}-\textless{}operator\textgreater{}-s\textless{}sample-number\textgreater{}-raman \textbf{Do not use plates with NaOH} (NaOH may form a precipitate with metal ions, interfering with Raman measurement) 
    \item After samples have been taken: place sample plate into the centrifuge for 5 min, 4 °C, max. rpm. 
    \item After centrifugation, transfer supernatant into last rows off the plate (turning the plate around so that supernatant of container 1 is in upper left corner after flipping the plate)
\end{enumerate}
\textbf{Version 1 (manual):} \begin{enumerate}
    \item  Pipet 50 uL of each sample into the sampling interface using a multichannel pipet
    \item  Start the measurement by executing the command in the terminal on the computer that runs the orchestrator (\textit{raman-client queue-measuring-tasks \texttt{--}grpc-channel 127.0.0.1:50051 \texttt{--}duration 10 \texttt{--}num-scans 2 \texttt{--}sample-volume 50 \texttt{--}clean-pump-volume 0.00125 \texttt{--}clean-flow-rate 0.0075 \texttt{--}pull-back-volume 0.000385 \texttt{--}pull-back-flow-rate 0.005 \texttt{--}sample-to-cuvette-pump-volume 0.000115 \texttt{--}sample-to-cuvette-flow-rate 0.0025 \texttt{--}move-sample-volume 2e-05 \texttt{--}sample-label \textless{}sample-label\textgreater \texttt{--}run-id \textless{}run-id\textgreater \texttt{--}number-of-wells 8 \texttt{--}device-id \textless{}device-id\textgreater \texttt{--}substrates-of -interest \textless{}substrates-of-interest\textgreater{}} More information on the command can be found in the \href{https://bvt-htbd.gitlab-pages.tu-berlin.de/kiwi/tf3/raman-orchestrator/cli.html#interface}{documentation}
    \item Discard the plate after sampling is done
\end{enumerate}
\textbf{Version 2 (automatic):}\begin{enumerate}
    \item Return the plate to its place on the Tecan deck
    \item Initiate measuring by setting isRamanSupernatant\_plate\_back.gwl to 1
    \item The Tecan will measure the samples automatically
    \item Discard the plate after sampling is done
\end{enumerate}
\textbf{Version 3 (offline Raman):}\begin{enumerate}
    \item After all the samplings are taken in the plate, and after centrifugation and transferring of the supernatant, freeze the plate at \SI{-21}{\celsius} until measurement
\end{enumerate} \\ 
\hline
Check spectrum quality & Periodically check the quality of the spectra being recorded to make sure that there is no clogging of the pipes or other dirt interfering with the measurements. For more details see part \hyperref[troubleshooting]{troubleshooting}\\ 
\hline
Regularly check liquid levels of canisters & An empty canister may lead to wrong measurements. \\ 
\hline
\end{tabular}
\end{table*}

\subsubsection{After a Run}\label{after-run}
In Table \ref{tab:after-run} the steps that need to be done after the run to shutdown the system are explained in more detail.

\begin{table*}[h]
\caption{After the run}
\label{tab:after-run}
\begin{tabular}{>{\raggedright\arraybackslash}p{0.15\linewidth}>{\raggedright\arraybackslash}p{0.75\linewidth}}
\textbf{Step} & \textbf{Details} \\ 
\hline
Flush the system with water & Flush the system at least twice to remove any residues of samples. The command you can use may look like this: \textit{raman-client queue-washing-tasks \texttt{--}grpc-channel 127.0.0.1:50051 \texttt{--}clean-pump-volume 5e-3 \texttt{--}clean-flow-rate 0.75e-2 \texttt{--}pull-back-pump-volume 3.85e-4 \texttt{--}pull-back-flow-rate 5e-3 \texttt{--}number-of-wells 8} (more information in the \href{https://bvt-htbd.gitlab-pages.tu-berlin.de/kiwi/tf3/raman-orchestrator/cli.html#interface}{documentation}) \\ 
\hline
Shut down the servers & Shut down the servers using Ctrl+c \\ 
\hline
Turn off the devices & Switch off the devices \\ 
\hline
\end{tabular}
\end{table*}

\subsection{Maintenance}\label{maintenance}
In Table \ref{tab:maintenance} the steps that need to be done periodically in order to maintain the system and to ensure a high spectrum quality in the long run are shown in more detail. In particular, we perform these steps before a 2mag cultivation.

\begin{table*}[h]
\caption{Maintenance}
\label{tab:maintenance}
\begin{tabular}{>{\raggedright\arraybackslash}p{0.15\linewidth}>{\raggedright\arraybackslash}p{0.45\linewidth}>{\raggedright\arraybackslash}p{0.275\linewidth}}
\textbf{What?} & \textbf{Details} & \textbf{When ?} \\ 
\hline
Deep clean & Use cotton swabs and isopropanol or ethanol to clean the pipetting interface, especially inside of the wells: 
\begin{enumerate}
    \item Flush the system with water
    \item Use cotton swabs and water to clean the interface
    \item Flush thoroughly with water to remove any isopropanol/ethanol traces
    \item Check spectrum quality
    \item Repeat if necessary
\end{enumerate} & before every run / when needed \\ 
\hline 
 Intense clean with acid & Clean the inside of the cuvette and tubes with acid:
 \begin{enumerate}
     \item Disconnect the tube to the system liquid container and insert the tube into acid (5\% hydrochloric acid)
     \item Flush the system multiple times to remove residues inside the tubes and cuvette
     \item Reconnect system liquid and flush thoroughly with water to remove remaining traces of acid
 \end{enumerate} & In case of dirtiness, especially inside the tubes or inside the cuvette after working with biological matter \\
 \hline
\end{tabular}
\end{table*}

\subsection{Troubleshooting}\label{troubleshooting}
In Table \ref{tab:troubleshooting}, we outline frequently encountered issues during the measurement process along with their solutions.

\begin{table*}[h]
\caption{Troubleshooting}
\label{tab:troubleshooting}
\begin{tabular}{>{\raggedright\arraybackslash}p{0.3\linewidth}>{\raggedright\arraybackslash}p{0.3\linewidth}>{\raggedright\arraybackslash}p{0.3\linewidth}}
\textbf{Issue} & \textbf{Possible causes} & \textbf{Possible fixes} \\ 
\hline
The intensity of the measured spectra is very low & There is no or not enough sample & Add sample \\ 
\cline{2-3}
& Shutter of Raman probe is closed & Open the shutter \\ 
\cline{2-3}
& The cuvette is not positioned correctly & Reposition the cuvette. There should be a spacer in the cuvette holder (with a small hole pointing upwards). If the spacer is in, press down the cuvette onto the spacer. The mirror should be visible through the cuvette. \\ 
\cline{2-3}
& There are a lot of cells in the sample & Centrifuge the sample and make sure to separate the supernatant and the pellet properly \\ 
\cline{2-3}
& There is another substance besides cells in the sample that increases the optical density of the sample. & Precipitates caused by e.g. NaOH can increase the optical density of the sample, resulting in low spectra intensity. Centrifuge again or try to remove interfering substances from your sample \\ 
\cline{2-3}
& There are a lot of air bubbles in the sample being measured & Check the way in which the sample is pipetted in (e.g. pipetting speed of liquid handling station). Check if all connections are tight. \\ 
\cline{2-3} & The cuvette is dirty or clogged & Clean the cuvette intensely with acid (see \hyperref[maintenance]{maintenance}) \\ 
\hline
There is no water coming out of the interface in the flushing step & The system water canister is empty & Refill the canister \\ 
\cline{2-3}
& The connections between the tubes are not tight & Retighten all the connections. The connection between the system water canister and the tubes is particularly prone to being untight \\ 
\hline
The waste water in the interface does not flow off & The waste canister is full & Empty the waste canister \\ 
\cline{2-3} 
& The waste tube has an air lock & Wiggle the tube to release the air or   disconnect the tube and connect it again. Check if connections are tight. \\
\hline
\end{tabular}
\end{table*}


\section{List of Material}
    We provide a list of material used for the microfluidic setup in Table \ref{tab:material}.
\begin{table*}[t]
    \caption{\textit{List of Materials}: To connect the devices with each other with tubes, we used the this microfluidic material.}
    \centering

    \begin{tabular}{>{\raggedright\arraybackslash}p{0.1\linewidth}>{\raggedright\arraybackslash}p{0.8\linewidth}}
    Amount & Article for Setup \\
    \hline
        10 & Nut, flangeless, PPS, 1/16'' OD, Headless, 1pc/PAK \\
    \hline
        2 & Flangeless Ferrule Tefzel (ETFE), 1/4-28 Flat-Bottom, for 1/16'' OD blue, 10pc/PAK \\
    \hline
        1 & Tubing, PTFE, 1/16 x 1.0 mm ID, 25 m/PAK \\
    \hline
        1 & Flangeless fitting PEEK 1/16in, 10 pc/PAK \\
    \hline
        1 & Flangeless Ferrule, for 2.0mm OD, 10 pc/PAK \\
    \hline
        2 & NS1D48042812 - CPC Plug 1/4-28 UNF Female Thread (Flat Bottom Port), with Shut-off Valve, EPDM Seal \\
    \hline
        2 & NS1D19042812 - CPC Coupling 1/4-28 UNF Female Thread (Flat Bottom Port), with Shut-off Valve, EPDM Seal \\
    \hline
        1 & PMCD4204 - CPC Coupling Plug 6.4 mm hose barb, panel mount, with shut-off valve, Buna-N \\
    \hline
        1 & PMCD1704 - Coupling 6.4 mm hose barb, with Shut-off Valve, Buna-N Seal \\
    \hline
        1 & Idex – Manufacturer No. P-387X – Article number GZ-02022-09 \\
    \hline
        1 & Idex – Manufacturer No. P-352X – Article number GZ-02021-95 \\
    \hline
        2 & Huenersdorf™ Wide-neck canister made of HDPE with UV protection \\
    \hline
        1 & Tefzel (ETFE) Tubing, natural, 1/16'' OD, 0.020'' ID, 15 m, 1 pc/Pkg \\
    \hline
        1 & Outer nut, flangeless, PEEK, short, headless, M6 Flat-bottom, for 1/16'' OD, 10 pc/Pkg \\
    \hline
        1 & Ferrule, flangeless, PEEK, 1/4-28 flat-bottom, 1/16'' OD, natural, 10 pc/Pkg \\
    \hline
        1 & Ferrule, flangeless, with stainless steel ring, 1/4-28 flat-bottom, for 1/16'' OD, natural, 10 pc/Pkg 
    \end{tabular}
\label{tab:material}    
\end{table*}



\end{document}